\documentstyle[prl,amsmath,aps,epsfig,twocolumn]{revtex}

\newcommand{\upd}{{\mathrm d}}
\newcommand{\eps}{\varepsilon}
\renewcommand{\phi}{\varphi}
\newcommand{\tf}{T_{\rm eff}}
\newcommand{\mtr}{m_{\rm tr}}
\newcommand{\tr}{t_{\rm rel}}

\begin{document}

\draft
\twocolumn[\hsize\textwidth\columnwidth\hsize\csname @twocolumnfalse\endcsname
\title{Shearing a Glassy Material: Numerical Tests of Nonequilibrium
Mode-Coupling Approaches and Experimental Proposals}

\author{Ludovic Berthier$^{1}$ and Jean-Louis Barrat$^2$}

\address{$^1$CECAM, ENS-Lyon, 46, All\'ee d'Italie, 69007 Lyon, France}

\address{$^2$D\'epartement de Physique des Mat\'eriaux,
UCB Lyon 1 and CNRS, 69622 Villeurbanne, France}

\date{\today}

\maketitle

\begin{abstract}
The predictions of a nonequilibrium schematic mode-coupling theory
developed to describe the nonlinear rheology of soft glassy
materials have been numerically tested in a sheared binary
Lennard-Jones mixture. The theory gives an excellent description
of the stress/temperature  `jamming' phase diagram of the system
both at the microscopic and the macroscopic levels.
In the present paper, we focus more particularly 
on the issue of an effective
temperature $\tf$ for the slow modes of the fluid, as defined from
a generalized fluctuation-dissipation theorem. As predicted
theoretically, many different observables are found to lead to the
same value of $\tf$, suggesting several experimental procedures to
measure $\tf$. New, simple experimental protocols to access $\tf$
from a generalized equipartition theorem are also proposed, and
one such experiment is numerically performed. These results give
strong support to the thermodynamic interpretation of $\tf$ and
make it experimentally accessible in a very direct way.
\end{abstract}

\vskip2pc]

\narrowtext

Glassy materials  are usually defined by the fact that their
relaxation time is larger than the experimental time scale. In
simple molecular systems, the associated glass transition
temperature corresponds to very high viscosities, making it
difficult to investigate experimentally their rheological
properties. In complex fluids (e.g. colloids, emulsions) it is
however possible to reach a glassy situation, in the sense of
large relaxation times, with systems having viscosities or shear
moduli that allow for rheological investigations~\cite{larson}.
Such materials have been described as `soft glassy
materials'~\cite{SGR}.

In its glassy state, a material is by
definition out of equilibrium. Physical properties are then a
function of the time $t_w$ spent in the glassy phase, a behavior
called aging~\cite{aging_review}. Interestingly,
recent experiments on various complex fluids 
have demonstrated striking 
similarities with other, more standard, glassy
systems~\cite{luca,cloitre,laurence,bonn,lequeux2}.
Upon imposing a   steady, homogeneous  shear flow, a
different kind of nonequilibrium situation is obtained. The flow,
characterized by the shear rate $\gamma$, creates a {\em
nonequilibrium steady state}, in which time translation invariance
is recovered~\cite{cukulepe}. This situation can therefore be used
to probe the glassy state, with the convenient feature of
having  the shear rate $\gamma$ rather than the waiting time $t_w$
as a control parameter. Moreover, this way of probing the
nonequilibrium properties of glassy systems is probably more
relevant experimentally than the aging approach, at least in the
case of soft glassy materials.

Recently, a general scenario was proposed for glassy systems
subject to an external forcing~\cite{BBK},
based on the study of mean-field models.
The rationale of this approach is that the equilibrium
dynamics of these models is equivalent to the `schematic'
mode-coupling approach of slowing down in supercooled
liquids~\cite{kithwo,gotze}.
The study of their nonequilibrium dynamics can thus be seen
as a {\it nonequilibrium} schematic mode-coupling
approach~\cite{aging_review,leticia3,cuku}.
To our knowledge, the mode-coupling theory of supercooled fluid
has not yet been extended to fluid under shear beyond the linear
response regime. However, in analogy to what was done for aging or
supercooled systems, it is sensible to bypass this aspect and to
carry out a direct comparison between mean-field predictions and
experimental or numerical results.

The aim of this work is thus to check on a realistic model
of a fluid the predictions that emerged from the theoretical
approach of Ref.~\cite{BBK}. Several earlier studies have been
devoted to simulating glassy sheared fluids. Ref.~\cite{Onuki}
focused on the dynamics at the molecular level. Our study is devoted
to more global aspects, with the aim of providing
experimentally testable predictions. Ref.~\cite{Liunagel} also
proposed to use the shear rate as a control parameter for
jamming systems.
Some dynamic aspects of a model similar to ours, with the
difference of being athermal (zero temperature) have been investigated,
as discussed in a companion paper~\cite{Liulanger}.

\begin{figure}
\begin{center}
\psfig{file=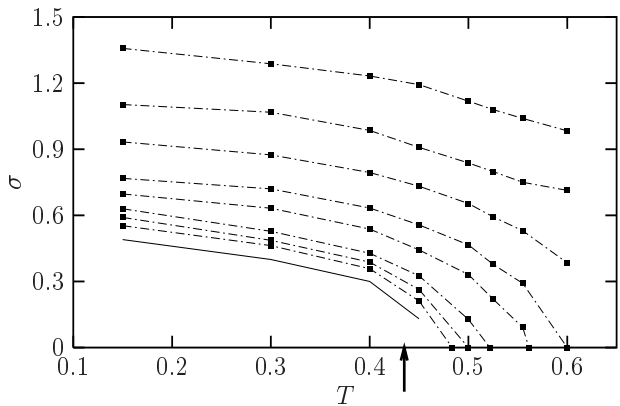,width=9.cm,height=6cm} \caption{The
$(\sigma,T)$ plane of the jamming phase diagram. The dashed curves
are the viscosity contour plots with $\eta=20$, 30, 50, 100, 200,
500, 1000 and 2000 (from top to bottom). The full line is the
yield stress $\sigma_0(T)$, from Eq.~(\ref{phenom}). Arrow marks
the mode-coupling temperature $T_c\simeq 0.435$.} \label{jamming}
\end{center}
\end{figure}

We have investigated the stress/temperature jamming phase
diagram (Fig.~\ref{jamming})
of the `standard model' for supercooled liquids,
namely a 3D 80:20 binary Lennard-Jones mixture,
in a simple shear flow defined by
$\boldsymbol{v} = \gamma y \boldsymbol{e_x}$.
The system consists of $N=2916$ particles in a cubic simulation box.
It has been characterized in much details at the reduced density $\rho=1.2$,
where we carry out our simulations, both in supercooled~\cite{kob}
and aging~\cite{barratkob} regimes.
Our simulations follow the protocol detailed in Ref.~\cite{BB}: the system
is first made stationary on a timescale of a few $\gamma^{-1}$. 
Then we perform our measurements in a range $T \in [0.15, 0.6]$ and
$\gamma \in [10^{-4}, 10^{-1}]$.
Standard Lennard-Jones units~\cite{kob,BB} are used. 

Following the theoretical analysis of Ref.~\cite{BBK}, our
approach consists in three main steps. We first investigate
macroscopic rheological aspects (flow curves). We then focus on
microscopic properties by an analysis of the density fluctuations:
structure factor, density-density correlation functions. Both
aspects will be discussed in much more detail in a longer
publication~\cite{BB2}. Last, we analyse the main new feature for
the field of rheology which emerged in Ref.~\cite{BBK}, namely the
 existence, behavior and properties of an effective temperature,
defined through a generalized fluctuation-dissipation
theorem~\cite{leticia3}. This point, together with its
experimental consequences are the main object of this paper.

The macroscopic behavior of the system under shear is
summarized in a stress/temperature phase diagram in
Fig.~\ref{jamming}. 
Our results are qualitatively similar to the one reported 
in Ref.~\cite{Onuki} on a different glass-former.
A Newtonian behavior is observed in the
high-$T$, low-$\sigma$ part of the phase diagram. 
This corresponds roughly to situations
where $\gamma^{-1}$  is larger than the relaxation time of the
fluid, which is thus only weakly affected by the flow. Outside
this region, we find that the viscosity of the fluid
decreases when the shear rate increases, a shear-thinning
behavior reported in various `soft glassy materials'~\cite{larson}.
We used the phenomenological relation~\cite{larson}
\begin{equation}
\sigma \simeq \sigma_0 + a \gamma^n.
\label{phenom}
\end{equation}
to extract the behavior of the
yield stress $\sigma_0(T)$, reported in
Fig.~\ref{jamming}. This `jamming transition' line was recently
experimentally investigated~\cite{weitz}. 
Although our results cover 3 decades in shear rate, we cannot report a
definitive functional form for the shear-thinning of the 
viscosity~\cite{footnote}
which can also be satisfactorily described by $\sigma_0 \equiv 0$,
which amounts to describe the system as a power-law fluid, $\eta
\sim \gamma^{n-1}$.
In the supercooled regime we find $n \simeq 1/3$, while at lower
temperatures, the exponent $n$ is temperature dependent with $n
\to 0$ when $T \to 0$, indicating that a yield stress could exist
in this limit only~\cite{BB2}. 
The latter power-law behavior is precisely the
one predicted  theoretically in Ref.~\cite{BBK}.

\begin{figure}
\begin{center}
\psfig{file=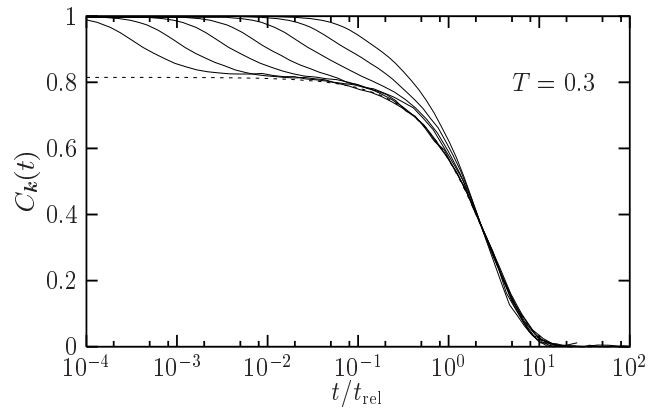,width=9.cm,height=6cm}
\caption{Correlation functions for $T=0.3<T_c$ and different values of
the shear rate, $\gamma=10^{-4}$, $3 \cdot 10^{-4}$, $10^{-3}$,
$3 \cdot 10^{-3}$, $10^{-2}$, $3 \cdot 10^{-2}$,
$10^{-1}$ (from left to right at $C=0.9$)
 can be collapsed if the time is rescaled by $t_{\rm rel}(\gamma)$.
The dashed line is a fit to a stretched exponential form, with
an  exponent $\beta = 0.95$.}
\label{corr3}
\end{center}
\end{figure}

Equilibrium mode-coupling theory gives rise to several
quantitative predictions regarding the scaling properties of the
intermediate scattering function,
\begin{equation}
C_{\boldsymbol{k}}(t) = 
\frac{1}{N} \sum_{j=1}^{N} \left\langle
\exp  \left(  i \boldsymbol{k} \cdot \big[ \boldsymbol{r_j}(t+t_0)-
\boldsymbol{r_j}(t_0) \big] \right)  \right\rangle,
\label{LEdef}
\end{equation}
when the glassy phase is approached
by lowering $T$ with $\sigma = 0$.
It was shown in Ref.~\cite{BBK} that similar scaling properties
are expected when the glassy phase is approached  by lowering
the shear stress $\sigma$ at constant
temperature (vertical line in Fig.~\ref{jamming}).
We have tested in a detailed way these predictions in
our simulation~\cite{BB2}.
Here, we only report the validity of the theoretically
predicted  `time-shear superposition property'~\cite{BBK}
in Fig.~\ref{corr3}, which shows that the slow decay of
$C_{\boldsymbol{k}}(t)$ has the scaling form
$C_{\boldsymbol{k}}(t) \simeq F(t/\tr)$, where the relaxation
time is defined, as usual, by $C_{\boldsymbol{k}}(\tr) \equiv e^{-1}$.
The scaling function $F(x)$ is well described by a stretched
exponential, $F(x) \sim \exp (-x^\beta)$, with an exponent $\beta$
which increases from $\beta \sim 0.77$ for $T \gtrsim T_c$
to the value $\beta = 1$ as $T \to 0$.

We now focus on the issue of an effective temperature for the slow modes
of the sheared fluid.
This quantity, naturally included in nonequilibrium mode-coupling
theories~\cite{BBK},
is defined through a nonequilibrium generalization of the
fluctuation-dissipation theorem (FDT)~\cite{leticia3}.
Consider two physical observables,
$O(t)$ and $O'(t)$, their connected cross-correlation function
$C_{OO'}(t) \equiv \langle O(t+t_0) O'(t_0) \rangle -
\langle O(t_0) \rangle \langle O'(t_0) \rangle$, and the
response function $R_{OO'}(t) \equiv \frac{\delta \langle O(t+t_0)
\rangle}{\delta h_{O'}(t_0)}$,
where $h_{O'}$ is the field thermodynamically conjugated
to $O'(t)$.
At equilibrium, both quantities satisfy the FDT,
$T R_{OO'}(t) = \frac{\upd C_{OO'}(t)}{\upd t}$.
The susceptibility
$\chi_{OO'}(t) \equiv \int_0^t \upd t' \, R_{OO'}(t')$ 
can be measured by applying a small (to ensure
linear response), constant field $h_{O'}$ between times $0$ and $t$ and
FDT implies a simple linear relation
$T \chi_{OO'}(t) = \big( C_{OO'}(0) -C_{OO'} (t) \big)$.
In the sheared fluid, an effective temperature is {\it defined} by
\begin{equation}
R_{OO'}(t) = -\frac{1}{\tf^{OO'}(C_{OO'})} \frac{\upd
C_{OO'}(t)}{\upd t} \label{FDTratio},
\end{equation}
where the $\tf^{OO'}(x)$ are \textit{a priori} arbitrary functions
of their argument, which may in general depend on the
observables $O$ and $O'$ under study.  For any pair of
observables, $\tf^{OO'}(x)$ can be measured in the sheared system,
by following the same linear response procedure as above, 
so that
$\chi_{OO'}(t) =
\int_{C_{OO'}(t)}^{C_{OO'}(0)}  \frac{\upd x}{\tf^{OO'} (x)}.$
The existence of an effective temperature is thus {\it
demonstrated} if a straight line is obtained in a
susceptibility-correlation plot parameterized by the time.
Obviously, the introduction of an effective temperature is of
`thermodynamic' interest only if this quantity
is actually independent of the
observables under consideration, $\tf^{OO'}
\equiv \tf$.
This crucial feature is true at the mean-field
level~\cite{leticia3,cuku},
and we shall prove that it is nicely satisfied in our model.

We have already shown the existence of such an effective temperature
for the slow modes of the system for a single correlation
function, Eq.~(\ref{LEdef}), at a given wavevector~\cite{BB}.
We also found that the essential phenomenological idea that
a system sheared more vigorously has a higher $\tf$
is indeed captured by this definition~\cite{BB}.

\begin{figure}
\begin{center}
\psfig{file=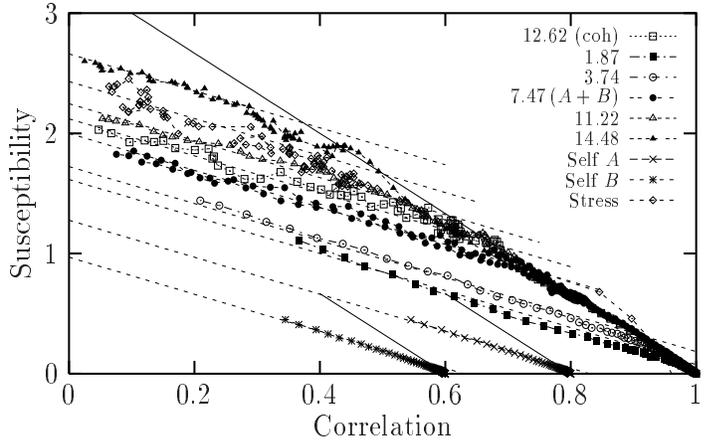,width=9.cm,height=6cm}
\caption{Ten representative susceptibility-correlation plots are shown
to be consistent with the same $\tf$ for the slow
modes (small value of the correlation).
Full lines are the equilibrium FDT of slope $-1/T$,
dashed lines have slope $-1/\tf$ with $\tf=0.65$.
Numbers refer to wavevectors (see also the text).}
\label{spec}
\end{center}
\end{figure}

Here, we go much further in our investigations and prove
that several different observables lead to the same value of $\tf$.
Since the numerical measurement of $\tf$ is very demanding,
our strategy has been to compute for $T=0.3$ and $\gamma=10^{-3}$
the value of $\tf$ from the observables of Ref.~\cite{BB}
with a great accuracy. We found $\tf \simeq 0.65$.
We then computed $\tf$ using different observables, and checked
that the value so obtained was compatible with $\tf=0.65$.
Our results are summarized in Fig.~\ref{spec} which shows 10 among the 
14 different susceptibility-correlation measurements we have performed.
All our data are well compatible with the single value of $\tf = 0.65$
for the slow modes of the fluid.

We first investigated the density fluctuations as
an observable, taking $O(t) = \frac{1}{N}
\sum_{j=1}^{N} \eps_j \exp \big( i \boldsymbol{k} \cdot
\boldsymbol{r_j}(t) \big)$, and $O'(t) = 2 \sum_{j=1}^{N} \eps_j
\cos \big( \boldsymbol{k} \cdot \boldsymbol{r_j}(t) \big)$. For
some wavevectors, this was done separately for both types ($A$ and
$B$) of Lennard-Jones particles (noted `$A+B$' in
Fig.~\ref{spec}). Taking $\eps_j = 1$ selects the coherent part of
the intermediate scattering function (noted `coh' in
Fig.~\ref{spec}), while $\eps_j = \pm 1$ selects the incoherent
one.
Such response-correlation plots could be obtained
experimentally in soft condensed matter systems. While the
correlation functions are reasonably easily obtainable through
light scattering experiments, the same is not true of response
functions. To obtain the latter, one has to
manipulate the particles through some externally applied
potential, modulated at the same wavevector as used in the light
scattering experiment. One suggestion would be to use some
non-index matched tracer particles in an index matched colloidal
suspension. The tracer particles would then be sensitive to the
intensity of the local electric field, as  in optical tweezers. An
interference pattern would actually realize the modulated external
potential considered here. Reading of the response could then be
obtained from the scattering at the wavelength corresponding to
this pattern.

We also used as a correlation the mean square displacement of a
tagged particle. The associated response function is the
displacement induced by applying a small, constant external force
to this tagged particle~\cite{parisi}. Both quantities are linked
by a FDT, the Einstein relation. We computed both quantities
separately for particles of type $A$ and $B$  (noted `Self $A$'
and `Self $B$' respectively). Its interest, especially in view of
experimental realizations, lies in the fact that the full time
dependence of the correlation and response functions is not needed
to extract $\tf$. Indeed, at large times, both quantities become
proportional to the time which defines the diffusion constant and
the mobility. One may therefore define $\tf$ simply as the ratio
of diffusion to mobility. Again, experiments could be considered
if tracer particles sensitive to an external force field (e.g.
magnetic particles) could be introduced into the system, and their
mobility measured together with their diffusion constant.

A completely different observable, relevant to flow
situations, is the stress $\sigma$. We have studied the case
in which the observables $O$ and $O'$ are equal to the
diagonal stress in the direction transverse to the flow,
$\sigma_{zz}(t)$. To add a field conjugated to $\sigma_{zz}$, a
compression $\delta L_z$ of the simulation box is realized by
rescaling all particle coordinates at time $t=0$, in analogy to
stress relaxation experiments. The corresponding curve is labelled
`Stress' in Fig.~\ref{spec}. Experimentally, the off-diagonal
component of the stress would be used
as the observable.  Preliminary results in this
direction, using an extremely sensitive rheometer, have been
obtained in aging colloidal systems~\cite{ludovic}.

\begin{figure}
\begin{center}
\psfig{file=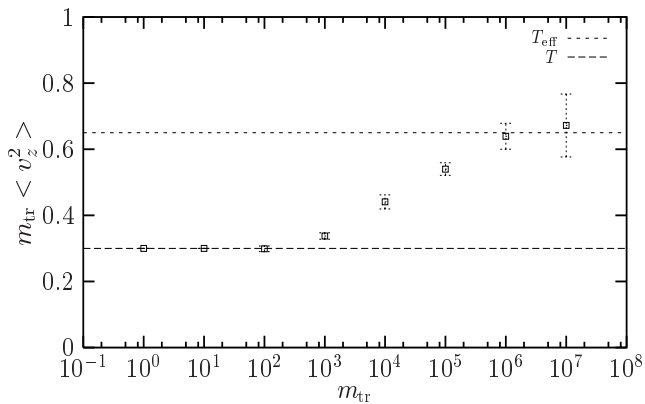,width=9.cm,height=6cm} \caption{Mass
dependence of the mean kinetic energy in the $z$ direction for
$T=0.3$ and $\gamma=10^{-3}$.
Horizontal lines are $T=0.3$ and $\tf=0.65$.
Error bars are evaluated from tracer to tracer fluctuations.}
\label{tracer2}
\end{center}
\end{figure}

Definition (\ref{FDTratio}) of $\tf$ implies
that, if the fluid is used as a thermal bath to equilibrate a
subsystem (a `thermometer')  of typical time scale $t_s \sim
\tr$,  the thermometer does not measure the microscopic
temperature, but rather the effective temperature associated with
its characteristic time scale~\cite{leticia3}. We propose here to
use tracers of mass $\mtr$ as a thermometer, since tuning $\mtr$
allows to control their vibration time scale $t_s$. We have
considered 10 massive tracers  with $\mtr \in [1,10^7]$, but being
otherwise identical to $A$ particles. Since $t_s \sim
\sqrt{\mtr}$, heaviest particles have a frequency typically $10^3$
times smaller than the light ones, implying $t_s \sim \tr$.
Reading of the temperature is done by measuring the average mean
square velocity of the tracers in the direction $z$. Results are
shown in Fig.~\ref{tracer2}, which shows that light particles
measure the bath temperature while heaviest particles
measure $\tf$.
This implies that a {\it generalized equipartition theorem} holds,
\begin{equation}
\left\langle \frac{1}{2} m_{\rm tr} v_z^2 \right\rangle =
\frac{1}{2} \tf.
\end{equation}
This result could be tested against experiments
involving for instance colloidal particles, the tracers,
in a  complex fluid, e.g. polymeric, or by
investigating rotational degrees
of freedom of non-spherical tracer particles.

This last result opens the way for new, simple determinations of
$\tf$ in out of equilibrium glassy materials. Indeed, no `complex'
dynamic functions such as correlations or susceptibilities are
needed here. In this context, it would be  interesting to
reproduce  Perrin's experiment on barometric
equilibrium of colloidal suspensions~\cite{perrin}. 
We expect indeed that the barometric
equilibrium of heavy particles inside a horizontally sheared fluid
should be ruled by $\tf$ instead of the room temperature.

All these results give strong support to the theoretical scenario
elaborated from mean-field theories to describe the rheology of
soft glassy materials~\cite{BBK}. Note in particular that $\tf$,
which we have shown to be a physically relevant quantity, is a
natural outcome of the theory. While numerical simulations provide
a test of the theory on short time scales,  systematic experiments
that could  test quantitatively existing theories of
nonequilibrium glassy dynamics are still needed. In that sense,
the situation is more or less similar to the mid-eighties when
schematic equilibrium mode-coupling were already derived, but with
little experimental confirmation of its main features. This is why
we tried, as much as possible, to suggest experimental
counterparts to our numerical measurements. We hope that our
findings and suggestions will motivate further experimental work
in the field.

We thank L. Bocquet, J.~Kurchan and W.~Kob for
discussions. This work was supported by the
PSMN at ENS Lyon and the CDCSP at Universit\'e de Lyon.


\begin{thebibliography}{199}
\bibitem{footnote}
Smaller shear rates, corresponding to larger time scales are presently
numerically unreachable.
\end{thebibliography}
\end{document}